# Do large EQs occur randomly in time? The Mexico EQ (20[th] of March, 2012, Mw = 7.4) as viewed in terms of local lithospheric oscillation due to M1 and K1 tidal components. A brief presentation.

**Thanassoulas[1], C., Klentos[2], V., Verveniotis, G.[3], Zymaris, N.[4]**

1. Retired from the Institute for Geology and Mineral Exploration (IGME), Geophysical Department, Athens, Greece.
   e-mail: thandin@otenet.gr - URL: www.earthquakeprediction.gr

2. Athens Water Supply & Sewerage Company (EYDAP),
   e-mail: klenvas@mycosmos.gr - URL: www.earthquakeprediction.gr

3. Ass. Director, Physics Teacher at 2[nd] Senior High School of Pyrgos, Greece.
   e-mail: gver36@otenet.gr - URL: www.earthquakeprediction.gr

4. Retired, Electronic Engineer.

## Abstract

The time of occurrence of the large EQ that occurred recently in Mexico (March 20[th], 2012, **Mw = 7.4**) is compared to the peak amplitude occurrence time of the local **M1** and **K1** tidal components. It is shown that the specific EQ occurred one (-1) day before the next following peak of the **M1** tidal component, and was delayed for only **+20** minutes after the corresponding **K1** tidal peak. Therefore, the specific seismic event complies quite well with the earlier proposed physical mechanism (lithospheric oscillation) that causes triggering of large EQs.

## 1. Introduction.

The aim of this very brief presentation is to show that the **M1** and **K1** tidal components play an important role concerning the time of occurrence of a large **EQ**. Actually, they provide the last decisive bit of stress load required in order to trigger a large EQ at an already critically stress charged seismogenic area. The physical mechanism that holds for the EQ triggering by the tidal waves has been presented in detail by Thanassoulas (2007) while specific examples have been presented from the Greek seismogenic area by Thanassoulas (2007), Thanassoulas and Klentos (2010), Thanassoulas et al. (2011), from New Zealand (Thanassoulas et al. 2011), from Japan (Thanassoulas et al. 2011a) and from a global analysis of the large seismicity (M ≥ 8) (Thanassoulas et al. 2011b).

In this brief presentation the time of occurrence of the recent very large EQ (Mw = 7.4) that occurred on 20[th] of March, 2012 in Mexico will be compared to the local seismogenic area tidal conditions within a time window of some days before and after the EQ occurrence time. The Rudman et al. (1977) method was used for the calculation of the corresponding tidal data.

## 2. The large EQ (Mw = 7.4R) of Mexico of the 20[th] of March, 2012.

The location of the large EQ is shown in the following figure (1) as presented by the EMSC.

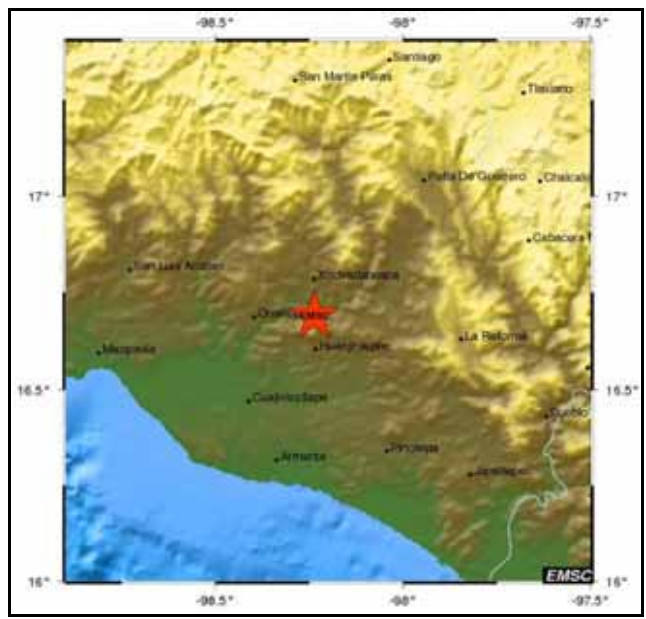

Fig. 1. Location (red star) of the large EQ (Mw = 7.4) of the 20[th] of March, 2012 in Mexico.

**The corresponding M1 (T = 14 days) component tidal data.**

The **M1** tidal oscillating component will be compared to the time of occurrence of the corresponding large EQ. The comparison is presented in the following figure (2).

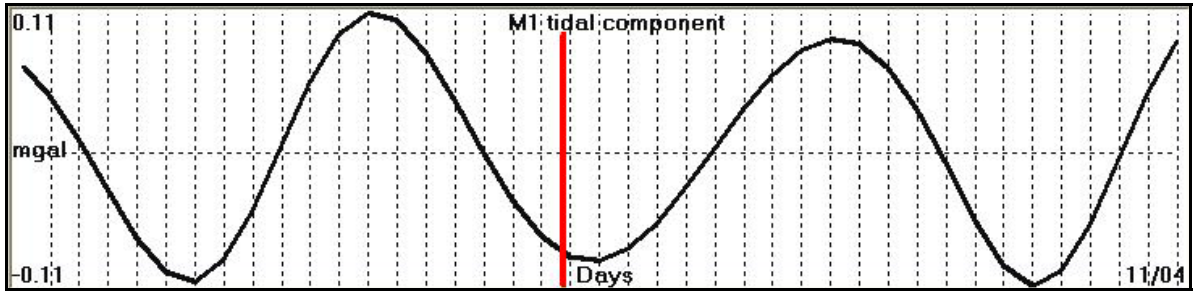

Fig. 2. Comparison of the **M1** tidal oscillation (black line, the lithosphere is forced to oscillate in the same mode) with the time of occurrence (red bar) of the EQ of 20th of March, 2012 (**Mw** = 7.4). Vertical scale is in mgals. The recording spans from: 20120302 to 20120410 (yyyymmdd).

In this case the EQ occurred, compared to the lithospheric tidal oscillation, one (-1) day before the peak of the **M1** tidal wave.

**The corresponding K1 (T = 24 hours) tidal data.**

Next, the lithospheric oscillating **K1** component will be compared to the time of occurrence of the corresponding large EQ. The comparison is presented in the following figure (3).

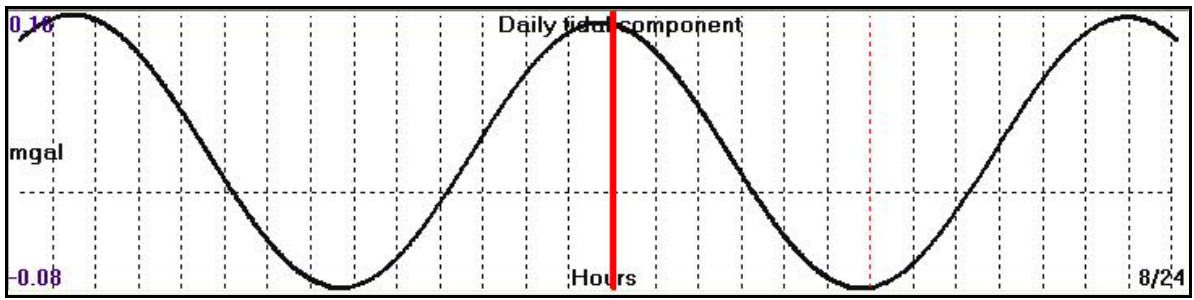

Fig. 3. Comparison of the **K1** tidal oscillation (black line, the lithosphere is forced to oscillate in the same mode) with the time of occurrence (red bar) of the EQ of the 20th of March, 2012 (**Mw** = 7.4). Vertical scale is in mgals. The recording spans from: 2012032004 to 2012032106 (yyyymmddhh).

The specific EQ deviates for only twenty (+20) minutes from the corresponding lithospheric oscillating tidal peak of the **K1** component.

### 3. Conclusions.

The inspection of figures (2) and (3) shows that:

a) the recent large EQ (**Mw** = 7.4) of Mexico occurred on the 20[th] of March of 2012 when the tidal component of **M1** achieves an oscillation amplitude peak (minimum) and consequently, the corresponding seismogenic area reaches a maximum of stress load on the very next day. Therefore the deviation (**dt**) of the time of the EQ occurrence from the **M1** tidal peak time is one (-1) day.

**dt = -1 day in terms of M1 tidal component.**

b) the recent large EQ (**Mw** = 7.4) of Mexico occurred on the 20[th] of March of 2012 and closely when the tidal component of **K1** achieves a daily oscillation amplitude peak (maximum) and consequently the corresponding seismogenic area reaches a same day short-term maximum of stress load. The deviation (**dt**) of the time of the EQ occurrence from the tidal peak time is determined from figure (3) as **dt = +20** minutes from the **K1** tidal peak time.

**dt = +20 minutes in terms of K1 tidal component.**

Consequently, the large EQ of Mexico that occurred on 20$^{th}$ of March of 2012 with a magnitude of $M_w = 7.4$ complied quite well in terms of its time of occurrence and the times of the peaks of the $M_1$ and $K_1$ tidal oscillating components with the physical mechanism earlier presented by Thanassoulas (2007).

The importance of the observed timing of the generation of large EQs, in relation to the M1 and K1 tidal peaks, is in the fact that it can be combined with simultaneous analysis of the earth's regional electric field and therefore a very short-term earthquake prediction can be utilized.

A detailed description of the postulated physical mechanism and more representative examples can be found in www.earthquakeprediction.gr

## 4. References.